\documentclass[reprint]{revtex4-2}
\usepackage{comment}
\usepackage{graphicx}
\usepackage{bm}
\usepackage{ulem}
\usepackage{color}
\usepackage{amsmath}
\usepackage{amssymb}
\usepackage{float}
\usepackage[space]{grffile}
\usepackage[dvipsnames]{xcolor}
\usepackage{braket}

\makeatletter
\renewcommand{\fnum@figure}{\textbf{Figure \thefigure}}
\makeatother

\definecolor{NQred}{HTML}{94636E}
\definecolor{NQsand}{HTML}{A79880}
\definecolor{NQgreen}{HTML}{6D8F84}
\definecolor{NQpurple}{HTML}{8883A1}

\begin{document}

\title{Hardware-Efficient Fault Tolerant Quantum Computing\\ with Bosonic Grid States in Superconducting Circuits}
\date{\today}
\author{Marc-Antoine Lemonde, Dany Lachance-Quirion, Guillaume Duclos-Cianci, Nicholas E. Frattini, Florian Hopfmueller, Chlo\'e Gauvin-Ndiaye, Julien Camirand-Lemyre, Philippe St-Jean}
\affiliation{Nord Quantique, Sherbrooke, Qu\'ebec, J1K 1B7, Canada}

\begin{abstract}
Quantum computing holds the promise of solving classically intractable problems. Enabling this requires scalable and hardware-efficient quantum processors with vanishing error rates.  This perspective manuscript describes how bosonic codes, particularly grid state encodings, offer a pathway to scalable fault-tolerant quantum computing in superconducting circuits. By leveraging the large Hilbert space of bosonic modes, quantum error correction can operate at the single physical unit level, therefore reducing drastically the hardware requirements to bring fault-tolerant quantum computing to scale. Going beyond the well-known Gottesman-Kitaev-Preskill (GKP) code, we discuss how using multiple bosonic modes to encode a single qubit offers increased protection against control errors and enhances its overall error-correcting capabilities. Given recent successful demonstrations of critical components of this architecture, we argue that it offers the shortest path to achieving fault tolerance in gate-based quantum computing processors with a MHz logical clock rate.
\end{abstract}

\maketitle

\section{Introduction}

Quantum computing is a revolutionary paradigm that reveals algorithmic capabilities beyond the reach of classical computers. Quantum computers are poised to have a transformative impact across diverse fields and industries, including material science, physics, cryptanalysis, machine learning, data science, finance, and optimization\,\cite{Jordan2024,Dalzell2023}. Driven by a growing community of developers venturing across all these disciplines, and now supported by increasingly sophisticated quantum software development kits\,\cite{Qiskit,Pennylane,Cirq,Bqskit}, the momentum in developing key algorithmic primitives and subroutines has drastically increased\,\cite{Arnault2024}. This progress is rapidly expanding quantum applications and moving the field toward practical implementations.

Algorithms offering super-polynomial speedups over classical computers present an optimistic view of the disruptive scientific and economic potential of quantum computing to enable solutions to classically intractable problems. Enabling practical implementations of such algorithms requires quantum computers of a certain size and with minimal errors per logical operation. While significant progress is being made in quantum computer performance across various modalities, achieving fault-tolerant operations with vanishing error rates remains crucial for implementing the majority of practically relevant applications\,\cite{Beverland2022}.

Recent resource estimations for Fault Tolerant Quantum Computing (FTQC) capable of executing tasks in machine learning\,\cite{Berry2024}, quantum chemistry\,\cite{Agrawal2024}, and finance\,\cite{Stamatopoulos2024} underscore the need for continued improvements and breakthroughs in hardware architecture design and performance to enable fast, reliable and scalable FTQC. Thus, quantum computing roadmaps focusing on fast clock-speed and fault-tolerant operations in resource-efficient hardware architectures are required to unlock the larger fraction of practical applications.

The foundational building block of FTQC is Quantum Error Correction (QEC) which provides the ability to detect and correct errors during quantum computations\,\cite{Gottesman2009,Girvin2023}. To overcome the inherent sensitivity of quantum systems to numerous noise sources, logical information in FTQC architectures is redundantly encoded, using various strategies, to prevent the unavoidable local physical errors from resulting in unrecoverable logical errors. These QEC techniques are crucial for reducing error rates (or errors per operation) from the current state-of-the-art $10^{-4}$ to the required $10^{-9}$ and below for practical applications\,\cite{Gidney2021,Lee2021,Goings2022}.

A key milestone towards successful QEC schemes is achieving the break-even point, beyond which the performance of an error-corrected logical qubit exceeds that of its best physical constituent; that is the point at which performing QEC overall improves the qubit logical coherence. The most widely adopted approach to QEC is to use an array of connected physical qubits to generate variants of the surface code\,\cite{Zhao2022, Krinner2022,Marques2022,GoogleQuantumAI2023,Sundaresan2023,Acharya2024}. A recent experimental surface code demonstration \cite{Acharya2024} needed around 50 physical qubits to provide the redundancy to reach the break-even point. Considering also the additional overhead required to perform quantum computation with logical qubits encoded in surface codes~\cite{Horsman2012, Litinski2019, Knill2004, Bravyi2005}, an overwhelming fraction of the total physical resources is therefore dedicated to the QEC portion of the algorithm being executed. This tremendous resources and engineering overhead highlights the need for a paradigm shift toward more hardware-efficient QEC techniques. 

\begin{figure*}[t]
    \centering
    \includegraphics[trim={0 0cm 0 0cm},clip, width=0.95\textwidth]{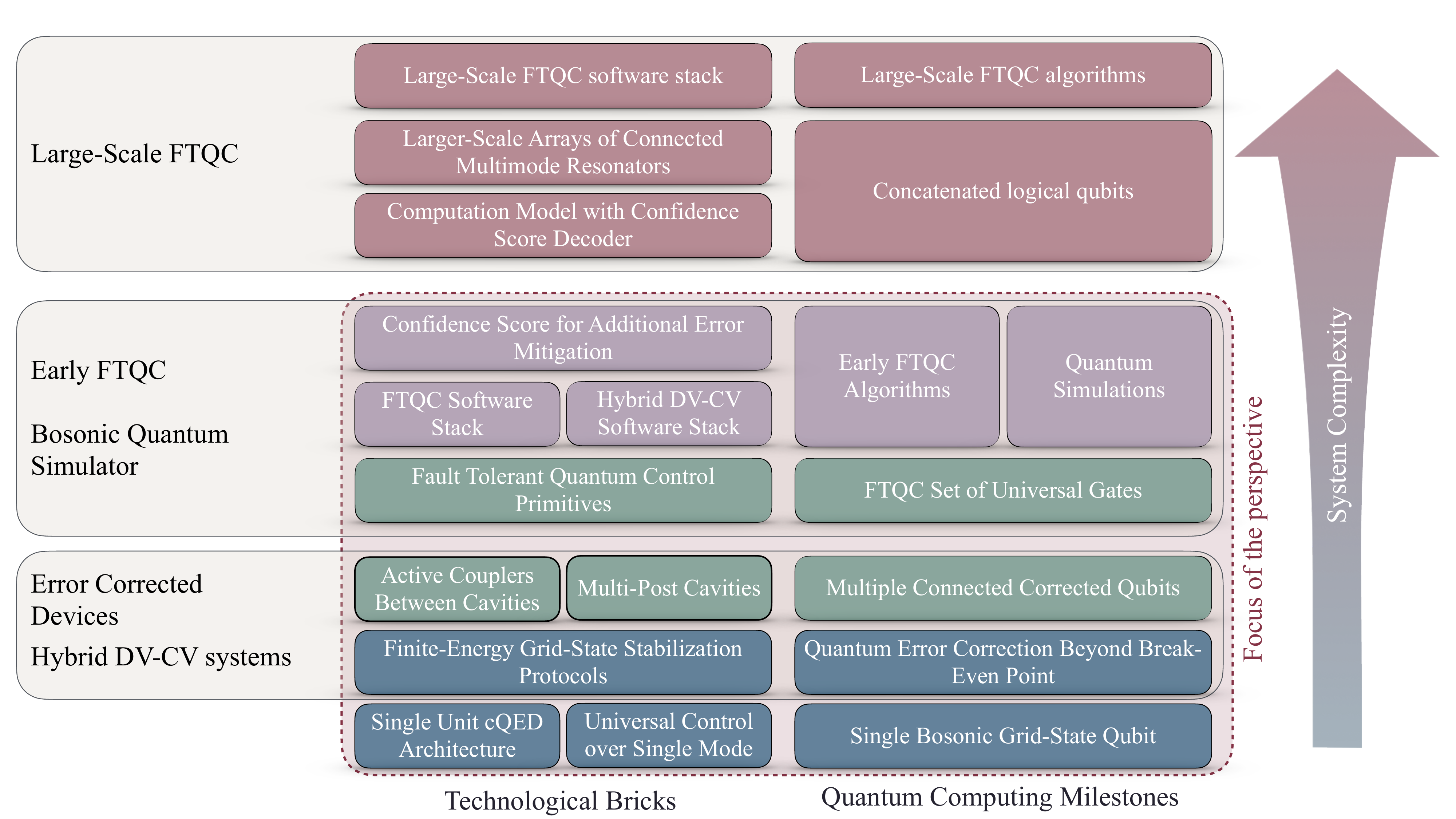}
    \caption{High-level roadmap towards FTQC with error-corrected bosonic grid-state qubits. The present perspective work focuses on the technological bricks, and corresponding quantum computing milestones, required to reach early FTQC, where each logical qubit is a multimode error-corrected bosonic code living in a single superconducting cavity. The FTQC Software Stacks refer to the Quantum Instruction Set Architecture (QISA) and the Transpiler/Compiler layer that separate a quantum algorithm from the associated control of the hardware. The ability to operate the bosonic quantum processor with an hybrid DV-CV computational model is also shown. The different key elements are expanded upon in the main text. The large-scale FTQC, where bosonic qubits could be concatenated with more standard qubit codes, such as qLDPC codes, is discussed as an outlook.}
    \label{fig:scope}
\end{figure*}

\begin{figure*}[t]
    \centering
    \includegraphics[trim={0 0cm 0 0cm},clip, width=0.99\textwidth]{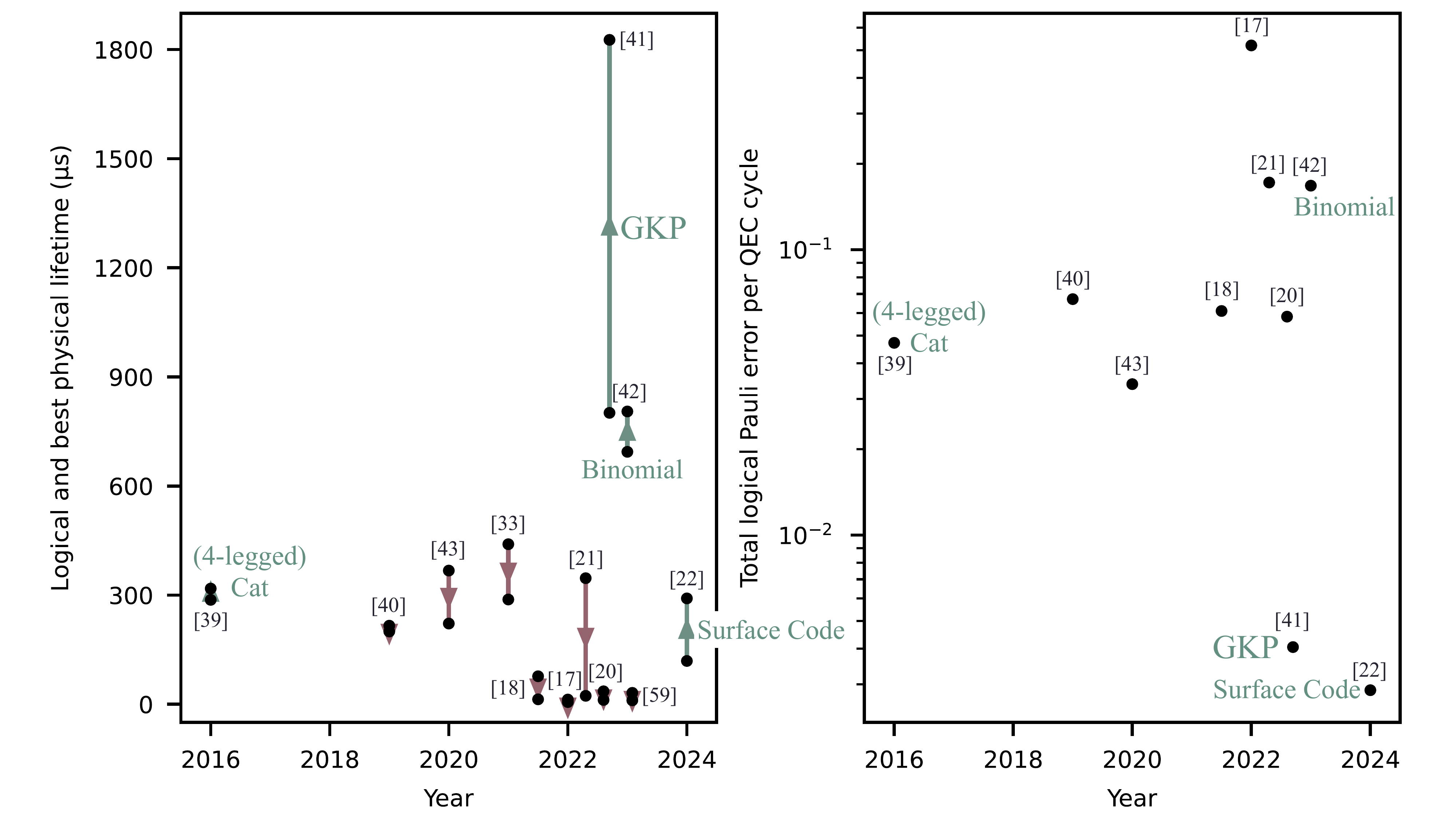}
    \caption{Review of the QEC attempts to go beyond the break even point in superconducting circuits. 
    In the left panel, the logical lifetime of the logical qubit after QEC is compared to the logical lifetime of the best physical constituent in the system. Green upward arrows (red downward) represent demonstrations beyond (below) the break even point.
    The right panel shows the total logical Pauli error per QEC cycle, i.e., the sum of the $X$, $Y$ and $Z$ Pauli error per cycle. Not shown in this figure, QEC demonstrations were also performed on other platforms such as trapped ions\,\cite{Ryan-Anderson2022} and cold atoms\,\cite{Bluvstein2024}, but none of those could go beyond the break-even point.}
    \label{fig:QEC}
\end{figure*}

A leading avenue devised to realize hardware-efficient FTQC is the use of bosonic encodings\,\cite{Cochrane1999, Gottesman2001,Mirrahimi2014,Michael2016,Albert2018, Terhal2020a, Gertler2021}. Unlike approaches relying exclusively on redundantly encoding across multiple physical systems, bosonic codes leverage the fact that the number of energy levels in quantum oscillators naturally extends far beyond the more standard two-level systems, in order to achieve logical protection. Having access to universal quantum control of such systems\,\cite{Krastanov2015,Heeres2015,Heeres2017,Eickbusch2022, Eriksson2024}, bosonic modes provide the ability to engineer superpositions of their energy levels into logical qubits that intrinsically have QEC properties, even at the level of a single quantum oscillator. Across modalities, very few attempts to QEC made it beyond the break-even point (see figure~\ref{fig:QEC}), and bosonic codes are the only strategy that can fulfill this using a single physical system; owning one of the best demonstration of quantum error correction across all quantum computing platforms\,\cite{Ofek2016,Hu2019,Sivak2023,Ni2023}.

Various bosonic encodings have been proposed to efficiently harness this larger dimension. Notably, grid states that rely on translational invariance\,\cite{Gottesman2001} have shown extremely competitive results for QEC\,\cite{Campagne-Ibarcq2019a,Sivak2023,Lachance-Quirion2024,Zheng2024}, making these a promising choice for building FTQC architectures\,\cite{Vuillot2019,Noh2020,Noh2022,Zhang2023,Lin2023}. Extending the concept beyond single quantum oscillators per logical qubit, bosonic codes based on grid states also provide a path to further reduce error rates in scaled-up, multi-mode cavities\,\cite{Royer2022,Conrad2022}.
 
Implemented in superconducting circuits, bosonic codes leverage two decades of engineering, thus providing a fast and scalable architecture for QEC. Not only do bosonic encodings in superconducting circuits allow hardware efficient QEC, but they also enable controls in the MHz regime, making them a promising prospect for implementing FTQC at scale for practical implementations.

In this perspective, we support the argument that a hardware-efficient platform built from bosonic grid states in superconducting circuits is the fastest path towards FTQC. Our focus in this manuscript is on mid-scale hardware architectures, demonstrating how errors can be autonomously corrected at the single physical unit level. We present techniques for executing fast logical gates and discuss how increasing the number of modes per unit can efficiently enhance device performance. We detail the key technological components required to achieve those essential quantum computing milestones toward that path. Beyond these fundamental building blocks, an integrated processing system will require a computational model at the architecture level to run actual algorithms. We briefly outline what such FTQC computing model would look like when using error-corrected bosonic codes. In addition, we highlight how the quantum processor architecture presented in this work can also be operated using an hybrid computational model, where discrete variables (DV) and continuous variables (CV) are combined in order to perform quantum simulations more efficiently compared to solely qubit-based computation\,\cite{Liu2024, Crane2024}. Altogether, this platform offers compelling advantages to implement early FTQC systems, that is, small-scale error-corrected processors capable of efficiently solving practically-relevant applications.

As an outlook, the technological components discussed here pave the way towards large-scale FTQC, where error-corrected bosonic qubits can be further concatenated with more standard qubit codes, such as quantum Low-Density Parity Check (qLDPC) codes. We expand how the unique ability to gather real-time confidence information about each error-corrected grid state can boost scaling efficiency towards large scale quantum computing architectures.
Taken together, these capabilities lay the groundwork for a QEC architecture based on bosonic multimode grid states, offering a pathway towards fast and fault-tolerant quantum computing in superconducting circuits.

\section{Bosonic Early FTQC Stack Overview}

As depicted in figure \ref{fig:scope}, the scope of this perspective manuscript focuses on early FTQC (eFTQC), where bosonic quantum processors can perform fault tolerant operations at a useful scale while avoiding the footprint and overhead of the hardware redundancy altogether. This is precisely what distinguishes mainstream attempts at fault-tolerance from the bosonic code approach: the ability of the latter to perform QEC at the most fundamental level of the stack, namely at the single unit level, in order to develop eFTQC processors. 

An eFTQC roadmap based on bosonic codes thus involves milestones that are different from the generic roadmaps towards FTQC. At the most fundamental level, it involves developing new types of individual units that can support those bosonic encodings, which in turn necessitate the development of universal control and readout of such encodings (figure \ref{fig:scope}, dark blue). Still at the hardware level, it also requires developing the architecture that allows the implementation of logical gates between these single-unit logical qubits (figure \ref{fig:scope}, olive). Given the fundamental differences of the bosonic approach, pushing the limits in terms of speed and precision of such universal control is as important as increasing the number of physical qubits on a chip in more standard strategies to FTQC.

At a higher level, it means devising the fault-tolerant quantum control primitives to be able to perform a set of universal FTQC gates (figure \ref{fig:scope}, sand).
Finally, it requires the development of a software stack adapted to this architecture towards the implementation of eFTQC algorithms (figure \ref{fig:scope}, purple).
In particular, the software stack supporting grid-state implementation needs to exploit efficiently the real-time confidence information that is gathered during the operations of each logical qubit. 

The longer view, where one could envision further concatenating the bosonic qubits with a more standard qLDPC qubit code for example (figure \ref{fig:scope}, red), lies outside the scope of the current paper and is only discussed as an outlook.

In what follows we describe in more details the necessary elements to build a hardware-efficient processor based on grid state encodings, starting from a more fundamental discussion about the richness of bosonic codes.

\section{Bosonic Codes as Logical Qubits}

The foundational building block on which this perspective work is based is the paradigm of bosonic encoding for implementing error-corrected logical qubits. In what follows, we present in more detail the fundamental concept behind their QEC properties. We start by briefly introducing different flavors of encodings to highlight the richness of this strategy. We then discuss in more detail the GKP code and its multimode extension, which outperforms other encodings on some critical metrics, and for this reason constitutes the main building block for the scalable QEC architecture discussed in this perspective paper.

\subsection{Logical Redundancy: Leveraging the Hilbert Space}

Standard physical qubit implementations in superconducting circuits, such as the transmon\,\cite{Koch2007}, in fact contain far more than two energy levels. The nonlinearity in their energy spectrum enables precise control over individual transitions -- therefore justifying their simplified treatment as two-level systems (TLS). When directly used as logical qubits, in addition to dephasing and relaxation events, any spurious population leakage outside the two controlled levels can become a source of logical error\,\cite{Nielsen2002}.\nocite{Li2024}

As such, the idea of expanding the information space by capitalizing on the readily available larger Hilbert space can help turn a liability into a strength: it can serve as a means to increase resources without requiring additional physical constituents and interfaces, along with their inevitable error channels. More precisely, a logical qubit encoded in $N$ energy levels of the Hilbert space of a single (bosonic) mode would adopt the following general form:
\begin{align} \label{eq:gen_enc}
    \ket{0}_L = \sum_{n=0}^N C_n^{(0)} \ket{n}, \quad \ket{1}_L = \sum_{n=0}^N C_n^{(1)} \ket{n}, 
\end{align}
where $\ket n$ is the $n^{\rm th}$ Fock state of the bosonic mode. From a simple counting argument, one can see that for $N > 2^M$, the dimension of the information space of a single physical mode exceeds the one spawned by $M$ TLS. These additional levels are the ones allowing for redundant encoding of the logical information with little additional hardware overhead.

In contrast, the standard paradigm for correcting physical errors involves redundantly encoding logical information in an interconnected array of TLS (referred to as physical qubits in what follows)\,\cite{Fowler2012}. For instance, one could encode a logical qubit using $M$ physical qubits in the following way:
\begin{align}
    \ket{0}_L = \ket{gg\dots g}, \quad \ket{1}_L = \ket{ee\dots e},
\end{align}
where $\ket{g}$ ($\ket{e}$) represents the lowest energy (first excited) state of a physical qubit. A local relaxation event, changing a single $\ket e$ into a $\ket g$, could be detected from a parity measurement and corrected through a majority vote\,\cite{Shor1995,Nielsen2002}.

In the standard paradigm, the hardware overhead to implement QEC protocols gets daring as the number of physical qubits increases. Not only does the amount of individually controllable constituents increases, leading to important engineering challenges in scaling the hardware and its control, but also additional noise channels are introduced alongside those constituents: this in turn contributes to increasing the demand on the QEC performance\,\cite{McEwen2021,Krinner2022,Marques2022,GoogleQuantumAI2023,Sundaresan2023,Acharya2024}. As shown in figure~\ref{fig:QEC}, all attempts to QEC using this strategy could not make it beyond the break-even point until very recently\,\cite{Acharya2024}.

\subsection{A Family of Bosonic Encodings}

The concept of accessing the vast Hilbert space inherent to each physical mode represents only a fraction of the story. For instance, one key question that must be addressed is how to best encode logical information robust to noise using $N$ energy levels per mode. Alongside the zoo of different bosonic encodings that have been proposed\,\cite{Jordan2024}, significant efforts have been put forward to compare the most promising strategies \,\cite{Albert2018}. Here, we briefly discuss some canonical examples illustrating the power of the bosonic code paradigm.

Arguably one of the most renowned examples of bosonic encoding is the (2-legged) cat qubit\,\cite{Cochrane1999, Mirrahimi2014}. In this code, the logical basis states can be defined as coherent states, which are by definition eigenstates of the annihilation operator and should thus be preserved under photon loss. However, any logical state that constitutes a coherent superposition of those basis states will be affected by photon loss, thus introducing computational errors. In more colloquial terms, while bit-flip errors are suppressed, phase-flips are not. Even if the cat qubit does not correct all errors on its own, it hints at the power of the bosonic encoding, where the use of a large Hilbert space allows to drastically change the characteristics of the noise channels. Importantly, such an asymmetry in the noise can be leveraged when concatenating cat qubits with qubit QEC codes, like the surface code\,\cite{Darmawan2021,Bonilla-Ataides2021,Chamberland2022} or qLDPC codes\,\cite{Ruiz2024}, to relax hardware resources requirements. 

One alternative encoding strategy that possesses the ability to correct up to $L$ excitation losses, $G$ excitation gains and $D$ dephasing events is the $(N, S)$ binomial code\,\cite{Michael2016}, where $S = L + G$ and $N = \max\{L, G, 2D \}$. The simplest example is the $(N=1, S=1)$ code where the logical states are
\begin{align}
    \ket{0}_L = \frac{\ket0 + \ket4}{\sqrt2}, \quad \ket{1}_L = \ket2.
\end{align}
Here, a single photon loss can be detected by measuring the change in the photon-number parity and then corrected by the application of a conditional unitary operation. While in this simplest example any error other than a single photon loss between two parity measurements will not be correctable, it already shows how the higher energy levels of the Hilbert space can be utilized to implement QEC protocols. 

\subsection{GKP Code}

\begin{figure*}[t]
    \centering
    \includegraphics[trim={0 0cm 0 0cm},clip, width=0.90\textwidth]{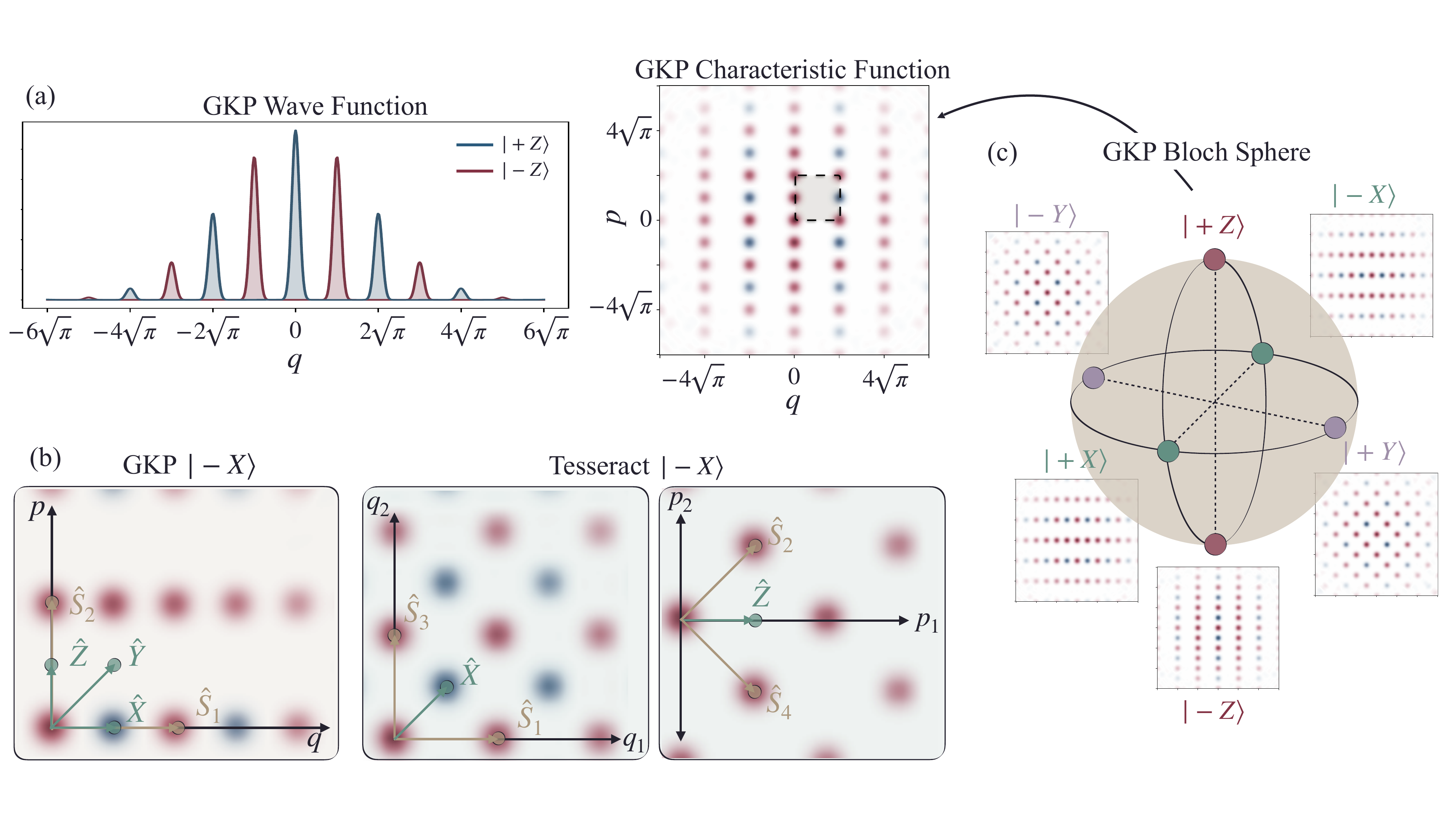}
    \caption{Translational invariant bosonic grid states. (a) Probability distribution in the position basis (squared norm of the wave function) of the $|\pm Z\rangle$ GKP states. Next to it is the characteristic function of $|+Z\rangle$ in the position $q$ and momentum $p$ quadratures (phase space) where red (blue) represents positive (negative) values on a scale relative to its maximal value. The shaded square represents the unit cell in which all the logical information is contained; highlighting the concept of logical redundancy. (b) Geometric representation of the Pauli logical operators and the code stabilizers for the GKP (left) in its two-dimensional phase space and Tesseract (right) codes in its four-dimensional phase space. The fact that for the Tesseract code, the logical operators and the stabilizers are not parallel is related to the {\it isthmus} property and is responsible for its increased robustness to control errors. The mathematical description of those operators is presented appendix \ref{app:stab}. (c) The logical Bloch sphere representing the GKP codespace, with the characteristic function of all cardinal states shown. All GKP states shown here have an average photon number of approximately 9, while the Tesseract code has the same amount per mode.}
    \label{fig:grid_states}
\end{figure*}

To find the shortest path to FTQC, a code should be chosen based on its ability to efficiently protect against the noise sources present in a physical implementation, while ensuring that the error correction procedure, as well as required gates, are physically implementable.

Given the same maximum average photon number used for encoding, the GKP code depicted in figure \ref{fig:grid_states} has been shown to outperform all other bosonic codes in\,\cite{Albert2018,Zheng2024}. The metric of comparison was the achievable fidelity from each code's optimal QEC protocol in presence of photon loss, which is the dominant noise channel in superconducting implementations. It has also been shown to maintain this out-performance when taking into account other relevant sources of error -- such as the dephasing channel -- under typical implementation conditions\,\cite{Leviant2022}. Of equal importance, the simple geometric structure of translation-invariant grid codes also allows the efficient implementation of error correction protocols, and logical gates\,\cite{Terhal2016,Campagne-Ibarcq2019a, Royer2020,deNeeve2022}.

The GKP code was originally constructed such that it has the ability to correct displacement errors in the position $q$ and momentum $p$ quadratures of the bosonic mode, as long as those are smaller than $\sqrt{\pi}/2$. Surprisingly, such ability translates into the ability to correct for several photon loss or gain errors with high accuracy\,\cite{Albert2018}. This contrasts for instance with the binomial code example from the previous section, which can exactly correct a single photon loss, but suffers a logical error for other errors.

This combination of both recovery outperformance (with respect to other bosonic encodings) and implementability explains its position as a natural choice to achieve FTQC in this platform.

\subsection{Going beyond single-mode encoding}

It is also possible to go beyond single-mode GKP codes by generalizing the concept of grid states and construct more sophisticated codes using multiple bosonic modes to encode a single logical qubit\,\cite{Gottesman2001, Royer2022,Lin2023,Wu2023}, therefore taking advantage of more robust information encoding in higher phase-space dimensions. 

One important motivation for the use of multi-mode grid codes lies in the fact that experimental techniques used to i) encode logical information in physical states, ii) recover said information when affected by a noise channel, and iii) ultimately decode back the information, are themselves error-prone processes that pose technical limitations. Multi-mode codes can be used to better address the single-mode GKP fundamental and technical limitations by providing an additional layer of robustness against these sources of error, as well as improve experimental accuracy and fault-tolerance of operations.

As an example, a widely proposed category of quantum codes for multiple bosonic modes is the concatenated codes\,\cite{Gottesman2001,Grimsmo2021}. Specifically, each mode encodes a single GKP qubit, which can then be used in a qubit stabilizer code capable of correcting arbitrary errors in up to $(d-1)/2$ (where $d$ is the code distance) GKP qubits. Concatenated codes could be particularly advantageous because they are robust to a wider range of errors, thus adding one layer of protection on top of the protection against small displacement errors in each mode. Such strategy is discussed in mode details in the outlook at the end of this manuscript.

The downside with concatenating GKP qubits with a stabilizer code is the hardware overhead that comes with their implementation. As illustrated in Figure \ref{fig:architecture} and discussed in section \ref{sec:multipost}, alternative hardware efficient strategies to scale the number of modes per qubit are accessible in superconducting circuits where resonators, with multiple and individually addressable modes, have been operated\,\cite{Koottandavida2024}. Such architecture allows finite scaling while still keeping the ratio of logical qubits over physical cavities to one. The ability to improve the error-correcting capability of a bosonic code by efficiently extending the number of modes per qubit is a crucial axis of scalability of the QEC architecture discussed in this perspective paper.

\subsection{The Tesseract Code: A First Step Toward More Robust Multimode Codes}
\label{sec:tesseract}

The Tesseract code constitutes the initial step of scaling the number of bosonic modes per qubit as it encodes a single logical qubit into two bosonic modes. It can be understood as the concatenation of two error-correcting rectangular single-mode GKP codes with the two-qubit error-detecting repetition code\,\cite{Royer2022,Lin2023}. 

As depicted in Figure \ref{fig:grid_states} and detailed in Appendix \ref{app:stab}, the geometrical structure of Tesseract and GKP codes differs in fundamental ways. For the GKP code, its stabilizers are parallel to its Pauli logical operators in the phase space. It turns out that such geometrical configuration allows specific control errors during the known QEC protocol implementation to translate into uncorrectable logical errors\,\cite{Royer2020} (see figure \ref{fig:sBs}). The extension of the phase space to four dimensions enables a choice of stabilizers such that logical operators and stabilizers are no longer parallel, thus improving the code robustness to control operations. This particularity is related to the so-called \textit{isthmus} property and is exploited by the Tesseract code.

Beyond the isthmus property, another fundamental advantage of the Tesseract code is discussed in Sec.~\ref{sec:confidence score} and illustrated in Figure \ref{fig:sBs}: the confidence information obtained during the QEC protocol is significantly more insightful compared to the GKP code\,\cite{Gauvin-Ndiaye2024}.

Those two fundamental differences highlight the richness that one can exploit by increasing the number of modes per logical qubit. The Tesseract code is the simplest example and going in even higher dimensions could unlock greater robustness to multiple sources of noise. 

\section{FTQC with Grid States in Superconducting Circuits}
\label{sec:controlling_large_Hilbert_space}

While identifying the optimal theoretical way of robustly encoding logical information in the large Hilbert space of bosonic modes is crucial, it is equally important to ensure that an efficient implementation of such code in a hardware platform is possible, as this is what ultimately defines its practical viability. 

\subsection{Building blocks in superconducting circuits}
\label{sec:native controls}

\begin{figure*}[t]
    \centering
    \includegraphics[trim={0 0cm 0 0cm},clip, width=0.90\textwidth]{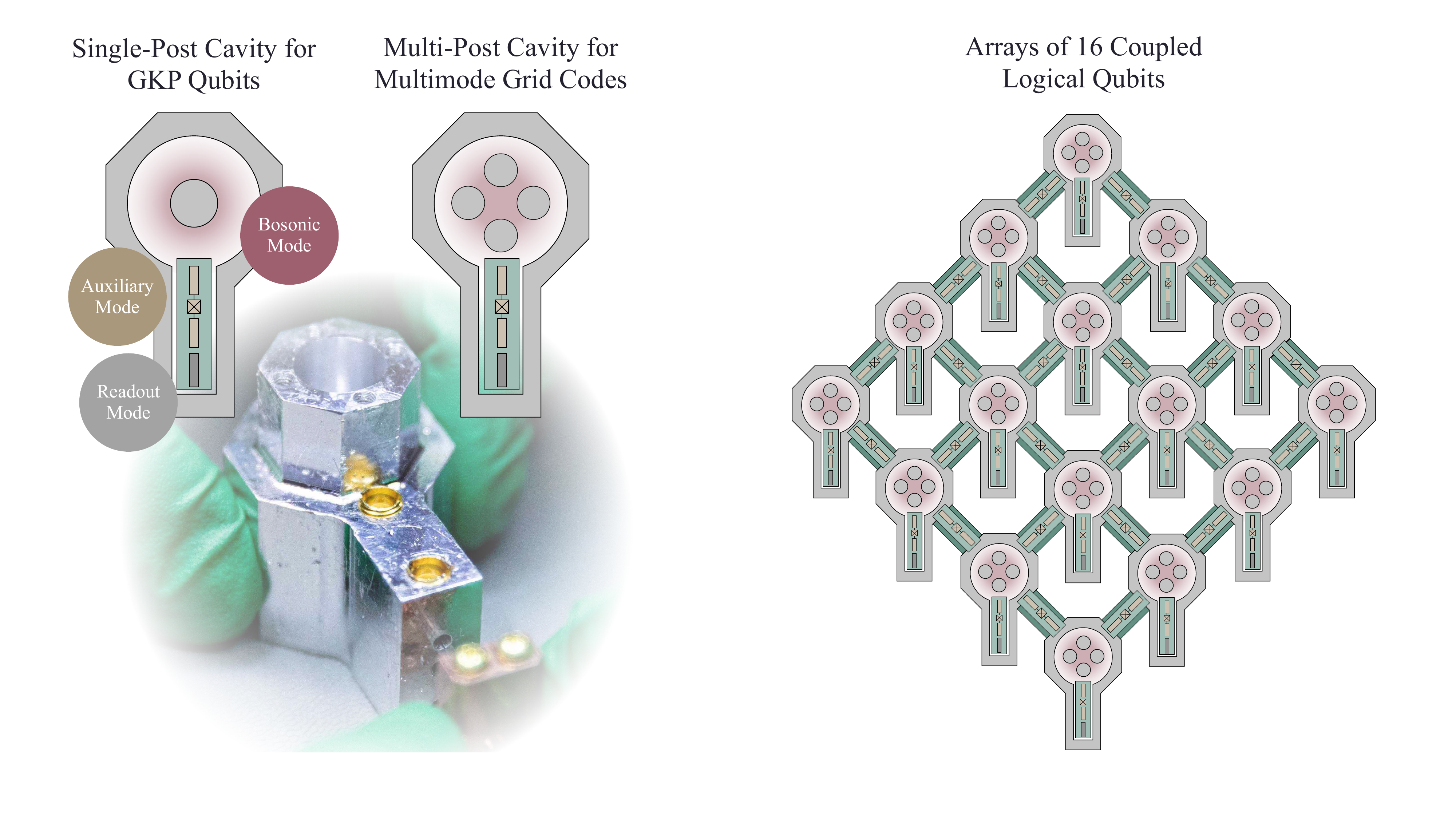}
    \caption{Hardware architecture for error-corrected quantum processor based on bosonic multimode grid states in 3D superconducting circuits. (Left) Single qubit architecture: A single-post (or multi-post) cavity houses the bosonic mode(s) in which the logical information is encoded and which is (are) weakly dispersively coupled with a non-linear auxiliary mode. A readout mode, composed of a lossy resonator, is also coupled to the auxiliary mode so that it can be readout and reset. The photo shows a high-purity aluminium three-dimensional superconducting cavity used in Ref.\,\cite{Lachance-Quirion2024}. (Right) Early FTQC architecture: 16 error-corrected bosonic qubits with nearest-neighbour interactions via active couplers are depicted. They can also be operated as an hybrid DV-CV processor as each cavity containing multiple bosonic modes is coupled to a nonlinear auxiliary element that can be operated as a physical qubit\,\cite{Liu2024, Crane2024}.}
    \label{fig:architecture}
\end{figure*}

Three key physical components are needed to operate a GKP qubit: (1) a long-lived harmonic mode (i.e.~a mode with a nearly linear energy spectrum, also referred to as an oscillator or a bosonic mode) wherein the logical information is encoded, (2) a coupled auxiliary nonlinear element used to address each level of the linear mode individually, and (3) an additional lossy resonator used to perform readout and reset of the auxiliary. The auxiliary element, with its readout resonator, is required for encoding, recovery, processing, and decoding of the logical information\,\cite{Ma2021,Cai2021,Grimsmo2021,Brady2024}. Such setup composes the {\it Single Unit cQED Architecture} box in Fig.\,\ref{fig:scope} and a three-dimensional implementation is shown in Fig.\,\ref{fig:architecture}.

For the long-lived harmonic mode, current state-of-the-art implementations rely on three-dimensional (3D) superconducting microwave cavities, such as seamless coaxial-stub cavities introduced in Ref.\,\cite{Reagor2016}. They are engineered to minimize seam losses with demonstrated lifetime that can exceed $25$\,ms with pure dephasing rate of less than $1$\,Hz\,\cite{Milul2023}. Significant advancements are being made in two-dimensional (2D) architectures\,\cite{Ganjam2024}, bringing us closer to the realization of high-quality on-chip bosonic code architectures.

Regarding the other two components, the simplest implementation comprises a transmon coupled to an on-chip superconducting resonator\,\cite{Blais2021,Axline2016}. This combination constitutes the most common superconducting physical qubit with lifetime that can exceed hundreds of microseconds\,\cite{Place2021,Kono2024} and for which all the key control primitives, which include arbitrary rotations, readout, and reset have been demonstrated\,\cite{Blais2021}.

Other nonlinear elements are being investigated for GKP qubit control, such as the fluxonium\,\cite{Manucharyan2009} or even other bosonic encodings such as cat qubits\,\cite{Leghtas2015a,Touzard2018,Lescanne2020,Grimm2020,Frattini2024,Reglade2024}, for which the natural bias of the noise channel could be exploited to achieve more robust control\,\cite{Puri2019a} with an early experimental demonstration\,\cite{Ding2024}.

The full toolbox of control operations required for a single GKP qubit is composed of single-element operations, plus an entangling operation between the oscillator and its auxiliary resource. The single-element operations are themselves composed of displacements of the oscillator state and arbitrary rotations, readout and reset of the auxiliary qubit. 

For the entangling operation, the echoed conditional displacement (ECD) is perhaps the most natural choice given that it is well-understood and can be realized experimentally,
all while providing universal control over the bosonic mode\,\cite{Campagne-Ibarcq2019a,Eickbusch2022}. The ECD entangling gate works particularly well in the regime of relatively weak auxiliary-oscillator dispersive interaction. Large effective entangling rates can be achieved despite small physical couplings, since the dispersive interaction can be effectively amplified by driving the linear cavity during the gate. Small physical couplings are advantageous since they lead to weaker unwanted nonlinearities and fewer losses of the oscillator inherited from the auxiliary\,\cite{Raimond2001,Eickbusch2022}. This set of operations constitutes the native {\it Universal Controls over a Bosonic Mode}, from which logical operations acting on a {\it Single Bosonic Grid-State Qubit} can be built, two key elements of the FTQC stack highlighted in Figure \ref{fig:scope}.

When it comes to connecting multiple cavity modes to perform multi-qubit controls or multimode code manipulations, many promising strategies have been demonstrated offering the option to operate in a wide range of parameter regimes. For example, a transmon dispersively coupled to distinct cavity modes can be either driven to engineer a direct generalized beam-splitter interaction via four-wave mixing\,\cite{Gao2018a,Burkhart2021}, or used to generate a selective multimode ECD if more than one cavity mode are being driven\,\cite{Lachance-Quirion2024b}. Similar to the standard ECD gate, the multimode ECD extends the universal control over multiple oscillator modes, making it a key building block for multimode codes such as the Tesseract code (see Sec.\,\ref{sec:tesseract}).

Another option is to use a superconducting nonlinear asymmetric inductive element (SNAIL)\,\cite{Frattini2017} to couple distinct cavities via three-wave mixing\,\cite{Chapman2023,Zhou2023}. Finally, state-of-the-art beam-splitter interaction between two microwave modes has been demonstrated using a differentially-driven DC-SQUID (superconducting quantum interference device), reaching gate fidelity exceeding $99.98\%$\,\cite{Lu2023}. Those different modalities are all viable options to implement {\it Active Couplers Between Cavities} to unlock {\it Multiple Connected Corrected Qubits} (cf. Figure \ref{fig:scope}).

\subsection{Efficient scaling to multimode codes}
\label{sec:multipost}

3D superconducting cavities are especially well-suited to scale the number of bosonic modes used to encode a single logical qubit: in addition to providing the required long-lived linear modes essential for leveraging bosonic encoding strategies, they offer a platform capable of efficiently implementing all-to-all coupling between multiple such modes. This is enabled by two key factors: i) the possibility of engineering 3D coaxial {\it Multi-Posts Cavities} that can house multiple long-lived linear modes (highlighted as a technological brick in figure \ref{fig:scope} and depicted in figure \ref{fig:architecture}), and ii) our ability to straightforwardly generalize the ECD gate to a multimode entangling gate in such a setting\,\cite{Lachance-Quirion2024b}. 

In these configurations, a single auxiliary can be dispersively coupled to all linear modes of a multipost cavity. By selectively driving multiple cavity modes simultaneously, an $N$-mode entangling gate can be performed in a time on the order of $\mu$s. This facilitates on-demand entangling gates between arbitrary modes, thereby enabling all-to-all active connectivity within a single multimode unit. Similar schemes can be implemented in 2D architectures.

The trade-off for this scalability is parallelism as any non-Gaussian control over one of the bosonic mode is performed via the auxiliary. However, by adding more auxiliaries to each unit, one could balance parallelism and hardware overhead, thus optimizing the system's overall performance.

While the number of modes per unit cannot be scaled arbitrarily, current technology allows for approximately $10$ modes per unit. This is akin to the situation in trapped ions, where the number of ions within a single trap is similarly limited\,\cite{Bruzewicz2019}. Access to an ensemble of around $10$ linear modes provides significant depth in terms of bosonic QEC strategy, offering a path for efficient scaling by partially alleviating locality constraints.

\subsection{Fast logical operations}
\label{sec:high frequency}

There are multiple additional advantages of working with superconducting circuits in the microwave regime. From a purely pragmatic standpoint, the large community working on such platforms all across the diversified field of quantum technologies -- as much in the academic sector as in the industry -- benefits from a rich worldwide ecosystem that drives technological breakthroughs. On a fundamental level, having access to strong nonlinear interactions between microwave-frequency modes offers deterministic universal control over the bosonic qubits.

More importantly, one crucial advantage is the timescale at which the set of native control operations can be implemented. For example, auxiliary rotations and oscillator displacements can both be performed in about $10$\,ns, while auxiliary readout and reset are usually performed in a few hundreds of nanoseconds within the circuit QED architecture\,\cite{Blais2021}. Regarding entangling operations, echoed conditional displacements are also usually performed in a few hundreds of nanoseconds,\cite{Campagne-Ibarcq2019a,Eickbusch2022,Lachance-Quirion2024}, while an excitation swap from a beam-splitter interaction can be performed in approximately $100$\,ns\,\cite{Lu2023,Chapman2023}. Since all key logical operations on GKP qubits (and their multimode extensions) can be generated from a shallow series of native controls, these operations can be performed on a microsecond timescale, leading to a megahertz computation clock rate.

\subsection{Encoding and decoding logical information in grid states}
\label{sec:encoding}

The first step for operating a bosonic qubit is preparing a known logical state into the large Hilbert space of an oscillator, also referred as the logical encoding. The last step is then the logical measurement, which can similarly be referred to as the decoding step. In the case of grid states in superconducting circuits, both of these steps are performed via the auxiliary nonlinear element\,\cite{Campagne-Ibarcq2019a,Grimsmo2021,Eickbusch2022,Kudra2022,Shaw2024}. As such, one can interpret the encoding step as copying the logical state from the auxiliary to the oscillator; similarly, decoding can be interpreted as mapping the logical information back from the oscillator to the auxiliary. Measurements, hence, are always performed on the auxiliary. When using a transmon as the auxiliary, the final step of the logical measurement involves the same standard procedure as for state-of-the-art superconducting industrial quantum processors\,\cite{Blais2021}.

Transducing an arbitrary bit of quantum information, from the oscillator to the auxiliary and vice-versa, can be performed by applying a series of alternating ECDs and arbitrary auxiliary rotations. In state-of-the-art demonstrations, circuit depths of no more than ten ECD gates have been used to prepare finite-energy GKP states\,\cite{Eickbusch2022,Lachance-Quirion2024}. For decoding, a single-shot quantum non-demolition (QND) measurement of the grid state can be performed using a shorter series of alternating ECDs and auxiliary rotations, followed by the measurement of the transmon along its $Z$ axis\,\cite{Gottesman2001,Royer2020,Shaw2024}. In the hypothetical case of an infinite-energy GKP qubit, the measurement of all logical Pauli operators and the two stabilizers [see Eqs.\,\eqref{eq:Stabilizer} and \eqref{eq:Pauli}] can be performed via a simple phase-space tomographic measurement that requires a single ECD gate, namely measuring specific points of the oscillator characteristic function\,\cite{Campagne-Ibarcq2019a,Fluhmann2020}. The measurement of the finite-energy logical Pauli operators and stabilizers is possible by extending the approach to a deeper circuit containing two ECDs\,\cite{Royer2020}. The generalization of the encoding and decoding techniques to multimode codes is straightforward given access to multimode ECD gates. 

\subsection{Quantum error correction}
\label{sec:Stabilization}

One of the key scientific breakthroughs that advanced GKP qubits from the theoretically most promising bosonic code to the leading candidate in experimental demonstration of QEC\,\cite{Sivak2023} is the invention of a {\it Finite-Energy Grid-State Stabilization Protocol}\,\cite{Campagne-Ibarcq2019a,Royer2020,deNeeve2022} (see figure \ref{fig:scope}). The idea is based on reservoir engineering, where a series of predefined controls are performed over the system to mimic the action of a fictitious cold bath capable of cooling the system to a particular ground state manifold\,\cite{Campagne-Ibarcq2019a,Royer2020,deNeeve2022}. The challenge is to engineer these controls so that the resultant steady-state corresponds to the GKP qubit manifold, that is, the set of states that constitute the Bloch sphere of the GKP qubit (see Figure \ref{fig:architecture}).

\begin{figure*}[t]
    \centering
    \includegraphics[trim={0 0cm 0 0cm},clip, width=0.90\textwidth]{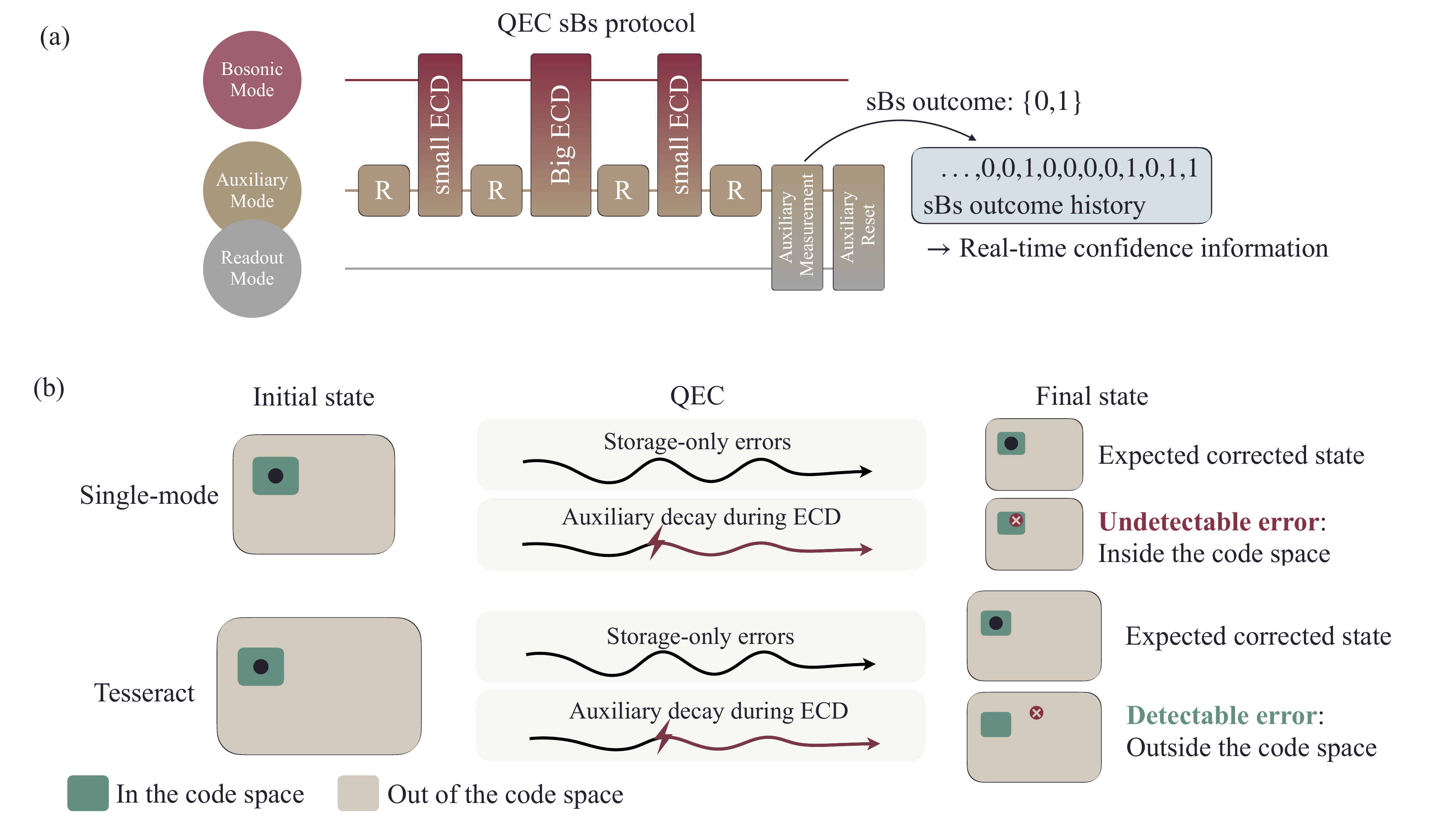}
    \caption{s\textbf{B}s QEC protocol with real-time confidence information. (a)\,Sequence of native controls to perform a single round of the s\textbf{B}s protocol composed of auxiliary rotations and ECD entangling gates. At the end of each s\textbf{B}s round, the state of the auxiliary is measured and then reset. The measurement results constitute the s\textbf{B}s outcomes and can be used to devise error mitigation strategies or boost the performance of an outer code decoder if the grid state qubits are concatenated with one. (b)\,Schematic representing the different impact of a control error on the s\textbf{B}s outcomes of the GKP compared with the Tesseract code. The worst case of an auxiliary decay happening halfway through the \textbf{Big} ECD shown in(a) can cause a logical error; it is silent for the single-mode GKP code and detectable with the Tesseract code.}
    \label{fig:sBs}
\end{figure*}

This approach led to the development of the well-known small-\textbf{Big}-small (s\textbf{B}s) protocol\,\cite{Royer2020,deNeeve2022} and related protocols. A notable accomplishment is efficiently executing reservoir engineering using the system’s inherent controls; its circuit implementation is illustrated in Fig.\,\ref{fig:sBs}. The s\textbf{B}s protocol has been used in one of the most successful experimental demonstration of {\it QEC Beyond Break-Even Point} to date\,\cite{Sivak2023}, surpassing the break-even point by a factor of 2.2.

Of note, an important feature of the implementation is that the state of the nonlinear auxiliary element is measured at the end of each QEC cycle. These collected measurements, referred as the s\textbf{B}s outcomes, can play a crucial role for subsequent information decoding steps. Hence access to these s\textbf{B}s outcomes bears the potential to be transformative for the QEC performance of GKP qubits; this is discussed further in Sec.\,\ref{sec:confidence score}.

\subsection{Logical gates with GKP codes}
\label{sec:gates}

While the QEC capabilities of GKP qubits are crucial, the complete picture also requires the ability to perform logical gates. 

The Pauli operations $\hat Z$ and $\hat X$ can be executed in software by redefining the Bloch sphere axis of the auxiliary qubit, a process known as a gauge update\,\cite{Royer2022}. Similarly, the Hadamard gate for GKP qubits can be performed by rotating the phase-space axis of the bosonic mode by $90$ degrees in software. These three key gates -- $\hat Z$, $\hat X$ and $\hat H$ -- can thus be applied virtually with perfect fidelity and almost instantaneously. Arbitrary single-GKP rotations, including non-Clifford gates, can be achieved through an auxiliary-mediated protocol\,\cite{Brady2024,Singh2024}. This protocol consists of a shallow sequence of alternating ECD gates and auxiliary rotations and is closely related to the s\textbf{B}s protocol. 

A key set of two-qubit logical operations, including, for example, the versatile exponential-SWAP gate\,\cite{Gao2018a}, can be implemented using an additional beam-splitter interaction\,\cite{Shaw2024} of the form $\hat H_{\rm BS} = g\hat a^\dag \hat b + g^*\hat b^\dag \hat a$, where $\hat a$ and $\hat b$ represents the destruction operators of the two coupled bosonic modes. 

A more complete set of two-qubit logical operations can be further implemented given access to the quadrature-quadrature interaction, which additionally includes the two-mode squeezing interaction $\hat H_{\rm TMS} = \eta\hat a \hat b + \eta^*\hat a^\dag\hat b^\dag$\,\cite{Shaw2024}. However, such an interaction does not conserve the number of excitations in the bosonic modes, thus leading to limitations especially for GKP qubits implemented using a small number of photons.

Similarly to single-GKP rotations, arbitrary two-qubit gates can also be achieved through an auxiliary-mediated protocol capable of implementing multimode ECD gates. While prone to auxiliary errors, this approach offers the possibility of implementing selective $N\geq2$ GKP entangling gates -- therefore offering a powerful avenue for extending to multi-qubit logical gates.

\subsection{Improving control robustness}

One of the main challenges with current hardware implementation of QEC in the proposed platform is that the encoding and decoding processes are not instantaneous and involve the transfer of quantum information to (or from) an unprotected auxiliary element; this in return limits the fidelity of the QEC protocol as it is subject to the finite coherence time of the auxiliary nonlinear mode. This is especially detrimental when the auxiliary undergoes a decay event, as not only the transfer process can be affected, but also because the ECD gate can potentially perform an unwanted logical operation, jeopardizing the logical information stored in the bosonic mode. Multiple pathways for improvement with regards to this vulnerabilty have been proposed, each targeting different levels of the stack. 

At the hardware level, employing noise-biased auxiliaries to minimize decay events remains a leading proposal for increasing the fidelity of code operations\,\cite{Puri2019a}. Another promising proposal towards improving robustness to auxiliary decay is to leverage higher-energy states of the auxiliary for error-transparent protocols\,\cite{Rosenblum2018,Reinhold2020,Ma2020,Ma2020b,Xu2024}.

At the control level, better-adapted pulse sequences derived from optimal control techniques can reduce the duration of each protocol and reduce the risk of problematic auxiliary decay\,\cite{Heeres2017}. Additionally, repeat-until-success and post-selected strategies involving multiple QND measurements can refine the preparation and measurement steps in the context of fault-tolerant quantum computation\,\cite{Reinhold2020}. 

One may also look for solutions to this auxiliary-decay problem at the level of the chosen code itself. For example, if the auxiliary decays at a critical moment during an s\textbf{B}s round (i.e.~near the halfway point of the Big ECD gate) while stabilizing the GKP code, it can result directly in a logical error that can be neither detected nor corrected. This is a consequence of having logical operators and stabilizers of the code which are parallel in phase space (see Appendix\,\ref{app:stab}). This once more motivates the extension to higher-dimensional phase space where it becomes possible to circumvent this vulnerability (see Sec.\,\ref{sec:tesseract}). 

Still from a control perspective, it should be noted that there are also other potential performance improvements to be made that are not related to auxiliary decay. The QEC protocols (e.g. s\textbf{B}s) that have been discussed and that are currently in use for these experiments are known to have limitations of their own. For instance it has been shown that the s\textbf{B}s protocol lies orders of magnitude away from being the theoretically optimal recovery strategy\,\cite{Zheng2024}. Consequently, at least from a theoretical perspective, we should expect that autonomous QEC protocols applied to GKP qubits and their multimode extensions have the potential to achieve performance levels much beyond what has already been demonstrated, which would bring fault-tolerant computation within reach.

The pathways for improvements mentioned here are not exhaustive. Designing and implementing these improvements in the underlying hardware, control sequences, code selection and operational strategies involves significant technological challenges and are the focus of ongoing research. Yet those avenues indicate that {\it Fault Tolerant Quantum Control Primitives} to implement a {\it Set of Universal FTQC Gates} in this platform is a realistic goal to achieve in a near future.

\section{Gaining Real-Time Confidence About Bosonic Qubits}
\label{sec:confidence score}

During the operation of a GKP quantum computer, mid-circuit measurements of the auxiliaries can provide real-time confidence about the GKP states, which can be used to significantly improve the performance of a fault-tolerant quantum computation.

The mechanism by which auxiliary measurements reveal confidence information can be understood as follows: physical errors, such as excitation loss, can cause the quantum state of a GKP qubit to leave the code space. Yet this doesn't necessarily mean that the logical information is irreversibly lost. Rather, the qubit state could reside in a different subspace with the same logical content. A schematic of this concept is presented in figure \ref{fig:sBs}.

We can thus understand the bosonic Hilbert space as a direct sum of multiple subspaces, known as error sectors\,\cite{Hopfmueller2024}. This distinction allows us to distinguish logical errors -- which occur within a given error sector and cannot be corrected -- from correctable errors that merely cause transitions between error sectors while preserving logical information.

When performing QEC, for instance by implementing the s\textbf{B}s protocol, measuring the auxiliary state after each round can yield additional information about error sector dynamics without revealing the logical state. As an example, a photon loss event will result with high probability in a s\textbf{B}s syndrome measurement outcome of "$1$" in a subsequent QEC round, indicating a transition back to the code space. It could, however, also be a signature that something is amiss with the state of the qubit, as could be the case if the state has completely left the logical codespace following an (uncorrectable) error.

Ideally, every logical error mechanism would always be accompanied by correctable error sector transitions, such that they would leave a detectable and unambiguous signature in the syndrome outcomes of subsequent QEC rounds. This logical error signature strongly depends on the code structure, which presents an opportunity to develop codes with this goal in mind.

For instance, the signature of displacement errors, which can be caused by auxiliary decay during an ECD gate, differs significantly between GKP and Tesseract qubits. A displacement error that would correspond to about half the length of the stabilizer of a GKP qubit leads to a logical error {\it within} the code space, and as such will go undetected by the s\textbf{B}s syndrome. In contrast, in the Tesseract code, a displacement error of exactly half that of a stabilizer, which could also result in a logical error, would at least be flagged by an s\textbf{B}s outcome "$1$" in subsequent rounds\,\cite{Gauvin-Ndiaye2024}. Interestingly, most logical errors happen to have a detectable signature in the Tesseract code QEC protocol, highlighting the advantage of using higher-dimensional Hilbert spaces for encoding logical qubits.

Therefore, the s\textbf{B}s outcomes offer confidence information about the reliability of each logical qubit at a given time. It is worth noting that real-time confidence information is available not only during QEC protocols, but also during other control operations, such as the encoding and decoding protocols described in Sec.~\ref{sec:encoding} and the auxiliary-mediated single- and two-qubit logical gates discussed in Sec.~\ref{sec:gates}.

In addition to the autonomous QEC performed during a computation, this real-time confidence information can help significantly boost the accuracy of an algorithm by giving access to {\it Additional Error Mitigation Strategies} during post measurements processing.
Error mitigation strategies have proven impactful in the absence of quantum error correction\,\cite{Kandala2019}, and could complement eFTQC computations where some errors could persist despite QEC protocols.

When grid codes are concatenated with an outer error-correcting code, the availability of time-resolved confidence information for each qubit could significantly boost the performance of the outer code\,\cite{Hopfmueller2024, Noh2020, Raveendran2022}. This potential improvement is reflected not only in the threshold of the outer code, but could also increase its effective distance, making a concatenated code more than the sum of its parts. Compared to ordinary qubits with the same error rate, this confidence information for GKP qubits could be used to achieve the same level of logical performance fewer qubits.

This approach can be compared to erasure qubits, a promising candidate for reducing the resource requirements for scalable QEC architectures\,\cite{Grassl1997}, which have recently been investigated in both neutral atom\,\cite{Wu2022,Ma2023, Sahay2023} and superconducting\,\cite{Kubica2023,Teoh2023,Levine2024,Chou2024,Koottandavida2024,deGraaf2024} systems.

Erasure qubits provide binary confidence information without revealing any logical information: a measurement outcome of "$1$" indicates that the logical information has been completely lost. In contrast, the full history of s\textbf{B}s outcomes can offer a more nuanced type of confidence information. This enables multiple strategies to enhance the reliability of quantum computation, including as a subset the erasure limit strategy -- i.e. of discarding any qubit as soon as it produces a single "$1$" outcome.

The availability of confidence information can be viewed as an example of error structure. Engineering and exploiting error structure is a major trend in quantum error correction, with other examples including erasure information and bias of Pauli errors. Confidence information combines aspects of both, since separate confidence information on distinct Pauli error channels is available, each associated with a specific stabilizer [see Eqs.\,\eqref{eq:Stabilizer} and \eqref{eq:Pauli} in Appendix \ref{app:stab}].

\section{Early FTQC Software Stack}

Once a quantum processor based on the architecture presented in the current perspective is developed, its usage will additionally require an operational strategy enabling the execution of meaningful quantum computations. Hence such a quantum processor must be equipped with a universal set of fault-tolerant logical gates or computational instructions, referred to as the Quantum Instruction Set Architecture (QISA).

It is expected that a user would typically want to use the quantum computer by designing high-level logical circuits corresponding to the algorithms they intend to run. However, these logical circuits may contain gates that are not included in the processor’s specific QISA, or involve operations between abstracted logical qubits that can't readily be mapped to physical hardware-level connectivity. A specialized program, known as a {\it compiler}, is responsible for synthesizing a QISA circuit that approximates the logical circuit with the desired accuracy.

Depending on the QEC strategy and the processor's architecture, compiling a logical circuit into a fault-tolerant computation can introduce significant overhead, drastically affecting the required computational resources. A canonical example of this is the well-studied surface code with lattice surgery~\cite{Horsman2012, Litinski2019}. In this approach, the quantum processor consists of a large array of physical qubits segmented into logical patches and routing auxiliaries. Logical operations are performed following a set of rules that require patch deformations, syndrome measurements, and, crucially, magic state distillation~\cite{Knill2004, Bravyi2005}.

A key advantage of the architecture presented in this manuscript is that the QEC overhead is significantly reduced, as each logical qubit corresponds to a single -- possibly multi-mode -- physical unit. QEC is naturally integrated into the instructions via control sequences, resulting in a much more manageable overhead. This contrasts sharply with conventional computational models designed to handle QEC strategies based on physical redundancy.

\section{Multiple Operational Modalities: Hybrid CV-DV Quantum Processor}

When defining a QISA and its associated compiler, it is important to note that the architecture depicted in figure~\ref{fig:architecture}, namely one based on inter-connected multimode cavities coupled to non-linear auxiliary elements providing universal control over the bosonic modes, can support multiple operational modalities: (error-corrected) qubit-based, analog-bosonic-based, or CV-DV hybrid quantum computing. The ability to operate directly on bosonic modes, avoiding the need to reduce everything to qubit-based simulations, can lead to substantial savings in hardware resources and enhanced performance. A recent comprehensive review systematically established the CV-DV QISA for this specific architecture~\cite{Liu2024}.

Implementations of some common algorithmic primitives for these systems have already been
developed, such as the CV-based Quantum Fourier Transform (QFT) and Quantum Random Walks.
Of note, the {\it non-Abelian QSP state transfer protocol} implementation of QFT described in\,\cite{Liu2024} relies on the same sequence of unitary operators as those used in the state preparation and stabilization protocols described earlier in Section~\ref{sec:controlling_large_Hilbert_space}. Hence the same architecture could readily be used in this context without requiring significant hardware design modifications.

Regarding quantum simulations per se, a common issue with DV-only quantum simulation is that it is typically limited to all-fermionic or all-bosonic systems. Additionally, the simulation of bosonic degrees of freedom on DV-only hardware is inefficient. Considering the relevance of families of problems involving both types of degrees of freedom, such as electronic-vibrational structure in quantum materials, the development of CV-DV quantum processors appears as a promising usage of the same hardware platform to unlock scientific and commercial value.
Recent work explores in details how such architecture can be used to efficiently simulate fermions, bosons, and gauge fields~\cite{Crane2024}.

\section{Outlook on Large-Scale Architectures: Concatenation With Outer Code}
\label{sec:LargeScale}

While $N$-mode bosonic codes implemented in a single $N$-post cavity might offer the best prospects for early fault-tolerant logical qubit implementations in the coming years, it is not clear that this approach can scale indefinitely within a single unit. However, these high-quality, error-corrected qubits with real-time confidence information, operating at performance levels orders of magnitude below threshold, can be concatenated into larger codes to achieve hardware-efficient larger-scale FTQC. 

One canonical approach is the surface code equipped with lattice surgery\,\cite{Horsman2012, Litinski2019}. A significant advantage of this scheme is that the confidence information from the s\textbf{B}s outcomes, is straightforward to leverage at the outer decoder level. For instance, in the now-standard matching decoder\,\cite{Higgott2023}, this confidence information can update the weights of the matching graph\,\cite{Higgott2023b}, potentially improving the threshold or the scaling of the sub-threshold residual error. Additionally, these strategies can be combined with recently developed post-selection methods that significantly improve the threshold with reasonable rejection probabilities\,\cite{Smith2024}.

The surface code lattice surgery scheme offers a remarkable range of space-time tradeoffs. However, when scaling to impactful computations, the qubit overhead remains prohibitive, even with highly reliable single-unit qubits, often requiring thousands or millions of qubits\,\cite{Beverland2022}. To address this issue, the community has explored the use of long-range interactions to enable denser, more efficient strategies, such as quantum qLDPC codes\,\cite{Bravyi2024}. While these require long-range interactions, they can be simulated with local interactions at a reasonable overhead\,\cite{Pattison2023, Berthusen2024}. 

Research on hardware-efficient FTQC remains very active. Some schemes propose applying lattice surgery directly to qLDPCs\,\cite{Cohen2022}, and ongoing discoveries in this area are likely to shape the future landscape. Preparing for what might come next involves actively exploring implementations of long-range interactions. This is further motivated by recent work on the so-called active volume of logical operations\,\cite{Litinski2022} which demonstrates that even limited long-range coupling can significantly reduce resource requirements when compiling logical circuits.

\section{Conclusion}

Bosonic encodings offer a promising solution to the scaling challenges inherent in traditional QEC strategies, which tend to rely heavily on increasing logical redundancy through the scaling of individually controlled physical two-level systems. Therefore, these bosonic encodings have the potential to achieve FTQC without the need to scale up to thousands of physical components per logical qubit.

Among the various bosonic code options, grid states, and specifically the GKP encoding, have demonstrated superior optimal error recovery channels \cite{Albert2018}. This optimality suggests that it is possible to achieve error rates low enough to enable FTQC under more reasonable physical requirements. Notably, while the GKP encoding has been demonstrated to comfortably surpass the break-even point~ \cite{Sivak2023}, early experimental protocols used in those demonstrations remain far from achieving the optimal recovery performance predicted by theory, meaning there is still substantial room for improvement in existing grid state QEC protocols.

Furthermore, generalizations of grid states to multi-mode encodings are expected to achieve higher performance than the simple single-mode GKP code. These more advanced encodings can help mitigate performance limitations related to the errors occurring in the required nonlinear auxiliary elements\,\cite{Royer2022}. Importantly, the enhanced richness of multi-mode grid state systems provides access to improved real-time confidence information about individual bosonic qubit gathered during the QEC process and gates, which is expected to help with additional error mitigation technique, while also significantly reduce requirements at the outer code level if there is a need for concatenation with more standard qLDPC codes in larger-scale FTQC architectures.

Taken together, these capabilities lay the groundwork for a hardware-efficient QEC architecture based on bosonic multimode grid states, offering a pathway toward early and large-scale fault-tolerant quantum computing in superconducting circuits with MHz clock rates.

\section{acknowledgements}

The authors would like to thank Alexandre Blais, Baptiste Royer and Christophe Jurczak for useful discussions and insights. The authors would also like to thank Volodymyr Sivak for providing the initial source code on which figure \ref{fig:QEC} is based. 

\appendix

\section{Technical Notes on Grid Codes}
\label{app:stab}

We here expand more on some mathematical details about the GKP and Tesseract codes structure in order to help understand the concepts of encodings with a finite mean number of photons as well as the {\it isthmus} property. 

The infinite-energy square GKP code is defined as the set of states that are joint eigenstates, with eigenvalue $1$, of the two commuting displacement operators
\begin{equation}
\label{eq:Stabilizer}
    \hat T_{q} = e^{il\hat q}, \quad \hat T_{p} = e^{-il\hat p}.
\end{equation}
Here $l=2\sqrt{\pi}$ is the dimensionless lattice constant.

Those operators are known as the code stabilizers (or generators) and when acting on a bosonic mode, they result in a translation in the phase space along the $p$ and $q$ quadratures respectively. The logical Pauli operators, also displacement operators, are given by
\begin{equation}
\label{eq:Pauli}
    \hat Z = e^{i\frac{l}{2}\hat q}, \quad \hat X = e^{-i\frac{l}{2}\hat p}.
\end{equation}
An important point is the fact that each stabilizer is parallel to one of the Pauli logical operator in phase space, as shown in figure \ref{fig:grid_states}. This means that if a control error happens during the measurement of a stabilizer, during the Big ECD of the s\textbf{B}s protocol for example, in a way that the associated translation is only partially performed, it can instead apply a logical operator and lead to a logical error.

The previous equations describe the idealized limit in which an infinite amount of photons would be required to generate such states. To ensure finite-energy states, required for experimental realization, each operator is dressed with an finite-energy envelope $\hat E_\Delta = e^{-\Delta^2 \hat a^\dag \hat a}$ such that $\hat T_{q/p,\Delta} = \hat E_\Delta \hat T_{q/p} \hat E_\Delta^{-1}$ (same for the logical Pauli operators). The mean number of photons used to encode a finite energy GKP code is roughly given by $\bar n \approx \frac{1}{2\Delta^2}-\frac{1}{2}$. The amplitudes of the ECD gates during the s\textbf{B}s protocol depends on $\Delta$. Therefore it is a parameter of the code that can be tuned to optimally correct the errors present in the experimental implementation.

In contrast to the GKP qubits, the Tesseract code has a geometrical structure where its stabilizers and Pauli logical operators are no longer parallel. More precisely, its infinite-energy stabilizers are defined as: 
\begin{align}
\hat{T}_1 = e^{-il\frac{\hat{p}_1}{2^{1/4}}} \quad \hat{T}_2 = e^{il\frac{\hat{q}_1+\hat{q}_2}{2^{3/4}}}, \\
\hat{T}_3 = e^{-il\frac{\hat{p}_2}{2^{1/4}}} \quad \hat{T}_4 = e^{il\frac{\hat{q}_1-\hat{q}_2}{2^{3/4}}}, 
\end{align}
with the logical operators given by
\begin{equation}
\hat{X} = e^{-il\frac{(\hat{p}_1+\hat{p}_2)}{2^{5/4}}} \quad \hat{Z} = e^{il\frac{\hat{q}_1}{2^{3/4}}}.
\end{equation}
As mentioned in the main text, this particularity is related to the so-called \textit{isthmus} property. The extension of the phase space to four dimensions enables this geometric arrangement, and it is this property that enhances robustness against control errors.

\bibliography{library}
\bibliographystyle{apsrev4-1}

\end{document}